\documentclass{emulateapj}
\usepackage{natbib}
\usepackage{epsfig}
\begin{document}
\title{KECK/DEIMOS Spectroscopy of a {\sl GALEX}  UV Selected
    Sample from the Medium Imaging Survey \footnote{Some of the
    data presented herein were obtained at the W.M. Keck
    Observatory, which is operated as a scientific partnership
    among the California Institute of Technology, the
    University of California and the National Aeronautics and
    Space Administration. The Observatory was made possible by
    the generous financial support of the W.M. Keck
    Foundation.}}

\author{ Ryan P. Mallery\altaffilmark{1}, R. Michael Rich\altaffilmark{1}, Samir Salim\altaffilmark{1}, Todd Small\altaffilmark{2}, Stephane Charlot\altaffilmark{3,4}, Mark Seibert\altaffilmark{2}, 
Ted Wyder\altaffilmark{2}, Tom A. Barlow\altaffilmark{2},
Karl Forster\altaffilmark{2},
Peter G. Friedman\altaffilmark{2},
D. Christopher Martin\altaffilmark{2},
Patrick Morrissey\altaffilmark{2},
Susan G. Neff\altaffilmark{5},
David Schiminovich\altaffilmark{6},
Luciana Bianchi\altaffilmark{7},
Jose Donas\altaffilmark{8},
Timothy M. Heckman\altaffilmark{9},
Young-Wook Lee\altaffilmark{10},
Barry F. Madore\altaffilmark{11},
Bruno Milliard\altaffilmark{8},
Alex S. Szalay\altaffilmark{9},
Barry Y. Welsh\altaffilmark{12}, 
Suk Young Yi\altaffilmark{10}}

\altaffiltext{1}{Department of Physics and Astronomy, University of California, Los Angeles, CA 90095-1562}
\altaffiltext{2}{California Institute of Technology,MC 405-47, 1200 East California Boulevard, Pasadena, CA 91125}
\altaffiltext{3}{Max-Planck Institut f\"ur Astrophysik, D-85748 Garching, Germany}
\altaffiltext{4}{Institut d'Astrophysique de Paris, CNRS, 98 bis boulevard Arago, F-75014 Paris, France}
\altaffiltext{5}{Laboratory for Astronomy and Solar Physics, NASA Goddard
Space Flight Center, Greenbelt, MD 20771}
\altaffiltext{6}{Department of Astronomy, Columbia University, New York, NY 10027}

\altaffiltext{7}{Center for Astrophysical Sciences, The Johns Hopkins
University, 3400 N. Charles St., Baltimore, MD 21218}

\altaffiltext{8}{Laboratoire d'Astrophysique de Marseille, BP 8, Traverse
du Siphon, 13376 Marseille Cedex 12, France}

\altaffiltext{9}{Department of Physics and Astronomy, The Johns Hopkins
University, Homewood Campus, Baltimore, MD 21218}

\altaffiltext{10}{Center for Space Astrophysics, Yonsei University, Seoul
120-749, Korea}

\altaffiltext{11}{Space Sciences Laboratory, University of California at
Berkeley, 601 Campbell Hall, Berkeley, CA 94720}

\altaffiltext{11}{Observatories of the Carnegie Institution of Washington,
813 Santa Barbara St., Pasadena, CA 91101}

\altaffiltext{12}{Space Sciences Laboratory, University of California at
Berkeley, 601 Campbell Hall, Berkeley, CA 94720}

\begin{abstract}
 We report results from a pilot program to obtain spectroscopy
 for objects detected in the {\it Galaxy Evolution Explorer} ({\sl GALEX}) Medium Imaging
 Survey (MIS). Our study examines the properties of galaxies
 detected by {\it GALEX} fainter than the  Sloan Digital Sky
 Survey (SDSS) spectroscopic survey. This is the first study
 to extend the techinques of  \citep{salim05}  to estimate
 stellar masses, star formation rates (SFR) and
 the b (star formation history) parameter for star-forming
 galaxies out to  $z\sim0.7$.
 We  obtain redshifts for 50 {\sl GALEX} MIS sources reaching $NUV=23.9$
 (AB mag), having counterparts in the SDSS Data Release 4 (DR4). Of our sample, 43 are starforming 
 galaxies with $z<0.7$, 3 have emission line ratios indicative
 of AGN with $z<0.7$, 
 and 4 objects with $z>1$ are QSOs, 3 of which are not previously cataloged.
We compare our sample to a much larger sample of $\sim$50,000  matched GALEX/SDSS galaxies 
with SDSS spectroscopy; while our survey is shallow, the optical counterparts to our sources
reach  $\sim$3 magnitudes fainter in SDSS {\it r} than the
SDSS spectroscopic sample. 
We use emission line diagnostics for the galaxies to determine that the sample
contains mostly star-forming galaxies. The galaxies in the sample populate
the blue sequence in the $NUV-r$ vs $M_r$ color-magnitude
diagram. The derived
 stellar masses of the galaxies range from 10$^8$ to 10$^{11}$
 M$_{\odot}$ and derived SFRs are between 10$^{-1}$ and 10$^{2}$
 M$_{\odot}$$yr^{-1}$. Our sample has
 SFRs, luminosities, and velocity dispersions that are similar
 to the samples of faint compact blue galaxies studied previously in
 the same redshift range by \citet{koo}, \citet{guzmana}, \&
 \citet{phillips}. However, our sample is $\sim2$ mag fainter in
 surface brightness than the compact blue galaxies. We find that the star-formation
 histories for a majority of the galaxies are consistent with
 a recent starburst within the last 100 Myr.
  \end{abstract}


\section{Introduction}
   Following the successful launch and early operations of the {\sl Galaxy
Evolution Explorer} satellite in 2003, we decided to undertake an initial
spectroscopic assay of sources detected by {\sl GALEX} using the DEIMOS
multiobject spectrograph \citep{faber03} on the Keck II telescope.  Although {\sl GALEX} data
have been matched to relatively deep optical surveys \citep{schiminovich} our study is the
first to extend the modeling technique
of \citet{salim05} from $z=0.25$ to $z\sim0.5$.

    The power of the {\sl GALEX} Medium Imaging Survey to select interesting objects is noteworthy.
We recall that the MIS observations are 1500 sec in duration,
imaging FUV and NUV simultaneously (\citet{martin} \& \citet{pm}).  
The {\sl GALEX} satellite has
investigated the ultraviolet universe to $z\sim1$ with the
goal of determining among other things, the star formation
rate (SFR) of galaxies in the local universe, $z\lesssim0.25$
\citep{salim05,martina,wyder,treyer} and the evolution of the
global star formation density out to $z\sim1$
\citep{schiminovich}. The ultimate goal of such work along
with other surveys and observations of galaxies at similar and
higher redshifts is to constrain the baryonic physics of
galaxy formation and evolution. Questions still remain as to
how the SFR depends on different factors such as environment,
mass, morphology etc. \citep{martin}. Previous work deriving SFRs
with a UV-selected sample was carried out by \citet{sullivan}
who derive SFRs both from UV luminosities at 2000\AA~and
H$\alpha$ luminosities from optical spectra for a sample of
galaxies at redshifts of $0<z<0.4$ detected in the UV by the
balloon-borne telescope FOCA \citep{milliard}. Estimates of
local star formation rates before this were only possible
through either optical emission line luminosities of the
recombination lines of Hydrogen, mostly H$\alpha$,  and
[OII]$\lambda$3727 or the far infrared luminosity from
10-100$\mu$m \citep{kennicutt}.

In this paper we present the observations and results from the pilot program for a
sample of objects detected by {\sl GALEX} and SDSS
with spectra obtained from the DEIMOS spectrograph at the Keck II telescope.
The analysis of this sample is enhanced greatly by the
techniques of Salim et al.(2005) that have been expanded to include the derivation of galaxy
physical parameters such as stellar mass, from {\sl GALEX} + SDSS photometry
and redshifts. The main goal of this paper is to characterize the
galaxies detected by {\sl GALEX} that are fainter than SDSS
Spectroscopic Galaxy Sample, in terms of 
mass, SFR,  and UV-optical color. 
We are predominantly (about two-thirds of the sample)
seeing the faint blue galaxy population \citep{ellis} and
are also sensitive to QSOs.  While {\sl GALEX}  observations have been undertaken
in many deep fields with much higher median redshifts
\citep{schiminovich} we have the advantage in this sample of obtaining spectroscopy
with Keck/DEIMOS, which gives wavelength coverage sufficient to derive
some detailed physical parameters from the spectra. 
 In \S 2 we describe the Sample and the photometric and
 spectroscopic data. Redshift and emission line flux
 measurements from the spectroscopy are presented in \S 3. The
 distributions of UV color, UV-optical color, and the SDSS
 {\it r} magnitude for the sample
 are described in \S 4. The derived galaxy parameters from SED
 fits are described in \S 5. We give a discussion and
 summary of our findings in \S 6 and 7. Throughout this paper we assume {\it H$_o$}=70
 km s$^{-1}$ Mpc$^{-1}$, $\Omega_{m}$=0.3, and
 $\Omega_{\Lambda}$=0.7.

\section{Sample and Observations}
\subsection{GALEX/SDSS Data}
  {\sl GALEX}  is a NASA Small
  Explorer Mission aimed to survey the UV emission from Galactic and extragalactic sources
  from 700km circular orbit
  \citep{martin,pm}. {\sl GALEX}  images the sky simultaneously in two bands, 
  the far-UV (FUV 1344-1786\AA) and the near-UV (NUV 1771-2831
  \AA). Each {\sl GALEX}  circular field is 1.25 deg. in  diameter. We use
  FUV and NUV magnitudes and magnitude errors
  derived in elliptical apertures \footnote{{\sl GALEX}  source
  detection and measurement is obtained from SExtractor
  \citep{bar}}. The photometry is taken from the {\sl GALEX}
  Internal Release 1.1 (IR1.1). The {\sl GALEX}  MIS photometry for
  tile number 10273  has a limiting magnitude, m$_{lim}$(AB)$= 23$. 
  
   We use optical photometry for our objects obtained from
   SDSS Data Release 4 (DR4) \citep{abaz}. 
   The SDSS photometric data are taken with
  the 2.5m telescope at Apache Point Observatory. Imaging is
  obtained in  {\it ugriz} bands
  \citep{fuk96,smith02}. The imaging data are photometrically
  \citep{hogg} and astrometrically \citep{pier} calibrated. An
  overview of the SDSS data pipelines and products can be
  found in \citet{stoughton}. 
 
\subsection{Sample}
Our sample is derived from objects with
  detections in the {\sl GALEX}  Medium Imaging Survey (MIS) tile 10273,
  centered at a right ascension/declination of
  17$^h$40$^m$32.07$^s$ $/$
  57$^o$10$^{\prime}$45.15$^{\prime\prime}$. 

The {\sl GALEX} field being much larger than the DEIMOS slit
mask, 
we selected objects from the {\sl GALEX}  field to populate
two DEIMOS slit masks.
 The two DEIMOS fields were initially chosen to maximize the number of
 MIS objects 
with optical counterparts in the SDSS classified as galaxies
with $FUV-NUV < 0$ photometry from an early version of the
{\sl GALEX}  data reduction pipeline. The number of galaxies
matching this criteria was much smaller than the number of
available slits,  and so to take full advantage of the DEIMOS
field of view, the slits  were populated by any galaxy detected by both {\sl
  GALEX} and SDSS.  In addition, the remainder of
available slits in both fields was populated with blue
stellar-like objects (QSOs and white dwarfs), main sequence
outlier stars ($u-g <1$ or $g-r<0$), and bright alignment
stars. The first DEIMOS field contains 26 objects classified
as galaxies in SDSS, and 10 blue stellar objects and main
sequence outlier stars. The second DEIMOS field contains 19
objects classified as galaxies in SDSS, and 9 blue stellar
objects and main sequence outlier stars.

\subsection{Spectroscopic observations} 
  The spectroscopic data were obtained at the Keck II
  telescope with the DEIMOS multi-object spectrometer on
  October 1, 2003 \citep{faber03}. Spectra were obtained using two slit masks,
  for a total 64 spectra. The 830 grooves mm$^{-1}$ grating
  was used with a slit width of $0^{\prime\prime}.73$  giving a
  resolution of $\sim2.5$\AA$~$FWHM. The signal-to-noise ratio
  obtained over the continuum of each spectrum varied between
  sources from 1 to  $\sim20$, depending on the brightness of the source.   
  The integration time for both slit masks was
  30 minutes. Flux calibration of the spectra was not performed
  as no flux standards were observed. 
 The spectra cover a wavelength range of $\sim
 5000-9000$\AA. Figure 1 shows a panel of several one
 dimensional spectra for star-forming galaxies  in the sample
 at a range of redshifts, plotted in the rest wavelength, and
 Figure 2 shows the spectra for objects with measured $z>1$. 

\section{Redshift determination and measurement of emission lines}
    In all, we measured 50 redshifts, 45 for sources
    classified as galaxies in SDSS and 5 for sources
    classified as stars in SDSS. Of those five, 
three are QSO's at $z>1$ and the other two are star-forming
galaxies. The remaining 14 of the sources classified as stars
by the SDSS have stellar spectra. The redshifts for objects
with detected emission lines were measured by eye. For objects
with  $z<1$, multiple emission lines were detected including
either H$\alpha$ and/or H$\beta$ and [OII]$\lambda$3727,
[OIII]$\lambda$4959 ,[OIII]$\lambda$5007,
[NII]$\lambda$6584,[SII]$\lambda$6717, and
[SII]$\lambda$6731. For all objects where the doublet
[OII]$\lambda$3727 was detected, H$\alpha$ was redshifted
beyond the wavelength range of the detector. For one source,
at z=1.028, we detected two features that we assigned as
[OII]$\lambda$3727 and MgII$\lambda$2800 emission. For the
other three sources with $z>1$ we find only 1 feature that we
identify as MgII$\lambda$2800, based on the width of the line
and the absence of other features in the spectrum. Figure 3 shows
the redshift distribution of the sample.  The mean redshift of the sample is $z_{mean}=0.421$. 
Two of the sources in the DEIMOS sample have corresponding
SDSS spectra. The redshifts obtained for these two objects
agree with the SDSS redshifts. One of the objects with SDSS
spectra is the QSO at $z=1.028$ and the other is a galaxy at
$z=0.176$. Both the DEIMOS spectra and the SDSS spectra for
the latter object have significant detections of only
H$\alpha$ and [NII]6584.

 We measure spectral emission line fluxes and equivalent
 widths for  [OII]$\lambda$3727, H$\beta$,
 [OIII]$\lambda$4959, [OIII]$\lambda$5007, H$\alpha$,
 [NII]$\lambda$6584, [SII]$\lambda$6717, and
 [SII]$\lambda$6731. Fluxes, equivalent  widths and errors for the emission
 lines were measured in IDL using the MPFIT function. The
 continuum of each spectrum was first fit by a polynomial,
 then the emission lines were simultaneously fit with
 gaussians. Table 1 lists the objects, their equivalent widths
 and flux ratios of  [OIII]$\lambda$5007/H$\beta$ and
 [NII]$\lambda$6584/H$\alpha$ for objects
 with a 3$\sigma$ measurement of at least one of the above
 lines. We did not perform a reddening correction to the
 fluxes, since the standard procedure requires either
 detections of both H$\alpha$ and H$\beta$ or a measurement of
 the radio continuum \citep{oster}. The effect of this on the flux ratios of  [OIII]$\lambda$5007/H$\beta$ and
 [NII]$\lambda$6584/H$\alpha$ is thought to be negligible due
 to the small wavelength separation of the emission lines. In the DEIMOS sample we
 only have detections of both H$\alpha$ and H$\beta$ for 18
 objects,  all at $z<0.3$, and chose for consistency not to
 perform corrections on any of the spectra. We note that the
 foreground Galactic extinction, E(B-V), calculated from the
 dust maps of \citet{schleg} is $\sim.05$ magnitudes for the field.

Fourteen objects have measured fluxes of H$\beta$,
[OIII]$\lambda$5007, H$\alpha$, and [NII]$\lambda$6584. Figure
4 shows the emission line diagnostic of \citet{bpt} used to
distinguish star-forming galaxies and Type II AGN. The solid
curve is the boundary between star-forming galaxies and AGN
determined through modeling of starburst galaxy spectra by
\citet{kew}. A star forming galaxy with only $\sim20$\% of the
optical emission line flux due to an AGN would lie above the
star-forming/AGN boundary \citep{kew}. The dashed curve is the
more stringent demarcation used by \citet{kauf03a} to distinguish
between AGN and star-forming galaxies in SDSS. The shaded
histogram shows the distribution of emission line ratios for a sample of ~51,000 SDSS/GALEX
objects (see \S 4). The curve of \citet{kauf03a}
distinguishes between the two different sources of emission
line flux in this large sample with the AGN occupying the
parameter space to the right of the distribution of star-forming
galaxies. In the DEIMOS sample only three galaxies of
the fourteen have emission line flux ratios above this curve,
implying that some portion of the flux is due to an AGN,
though it is still within the errors that the emission line
flux for two of these objects results entirely from star-forming regions inside the
host galaxies.   Objects can also be AGN if $\log$
[OIII]/H$\beta>1$ or $\log$ [NII]/H$\alpha >0.3$. Another 13
objects also have only detections of H$\beta$ and
[OIII]$\lambda$5007. Of the thirteen, one has a flux ratio of
$\log$ [OIII]/H$\beta>1$, indicative of an AGN. \citet{sullivan}
find a similar ratio of galaxies classified as
star-forming/AGN from examination of emission lines in their
sample of UV selected galaxies. 

\section{Magnitude and color distribution}
  The DEIMOS sample is $\sim3$ magnitudes deeper in {\it r}
  than the SDSS spectroscopic sample. Figure 5 shows a
  histogram of the  SDSS {\it r} magnitudes
for the sample plotted with the SDSS {\it
  r} magnitude for {\sl GALEX}  MIS sources in IR1.1 with
matches in the SDSS DR2 spectroscopic sample. The IR1.1/DR2
sample consists of $ \sim51,000$ galaxies at redshifts
$0.005<z<0.25$ with derived star-formation histories (SFHs) by
SED fitting (see \S 5). The IR1.1/DR2 histogram is normalized
to the size of the DEIMOS galaxy sample. The IR1.1/DR2 sample
is mostly bounded in {\it r} by the SDSS spectroscopic survey
limits, ${\it r}<17.7$ \citep{strauss}.

  Using the redshifts determined for the DEIMOS sample we
  calculate the absolute {\it r} magnitude, M$_{r}$, of
  galaxies in the sample, k-corrected to the mean redshift of
  the IR1.1/DR2 sample, $z=0.1$. K-corrections for all
  bandpasses were calculated using the publicly available code
  of \citet{blanton} version 4\_1\_4.
   In Figure 6 we construct a
   color magnitude diagram (CMD) of $NUV-{\it r}$ versus
   $M_{r}$. We overplot
   our sample onto the IR1.1/DR2 sample, plotted as a shaded
   contour plot.  In the figure the diamond symbols correspond
   to DEIMOS objects with spectroscopically measured redshifts
   at $z<0.25$, and the crosses correspond to objects with
   spectroscopically measured redshifts at $z>0.25$. The contours
   enclose 40\% and 80\% of the objects in the  IR1.1/DR2 sample.

The CMD of the IR1.1/DR2 sample clearly shows the bimodality
of galaxies seen by {\sl GALEX} in the nearby universe,
$z<0.25$, with distribution peaks at blue $NUV-{\it r}$ colors
of $\sim 3$ and at red $NUV-r$ colors of $\sim 6$
\citep{wyder}. We henceforth call these the blue and red
sequences. The galaxies in the blue sequence are generally
late-type in morphology and have spectra with emission
line-ratios indicative of star-formation
\citep{salim05,jarle}, while the galaxies in the red sequence
typically have absorption line spectra and lower
star-formation rates than blue sequence galaxies. The majority of
the objects in the DEIMOS sample lie in the region of the
diagram occupied by blue star-forming galaxies in the
IR1.1/DR2 sample, mostly along the blue edge of the
distribution, with only one having a $NUV-r$ color greater
than 4.  The spectrum for this object, at $z=0.077$, is shown
in Figure 1. The spectrum shows absorption features: NaD, and
MgIb, but also shows weak emission features of H$\alpha$,
[NII]6584, and [SII]6717,6731.

In the bottom-left panel of Figure 7 we again plot the color
magnitude diagram for the sample, highlighting the objects
with $FUV$ detections (filled circles).
The other 3 panels of this figure show the UV color
distribution for the 24 galaxies in the DEIMOS sample with
$FUV$ detections. Plotted for reference are galaxies from the
IR1.1/DR2 sample with $FUV$ detections shown as the shaded
countour plots  or shaded histogram in the respective
panels. Besides the 3 galaxies in the DEIMOS sample with
$FUV-NUV <0$ the UV color for this sample shows a similar
distribution with the IR1.1/DR2 sample. In the FUV CMD the
objects with M$_{r}<20$ show a range of UV color not seen in
the local blue sequence galaxies of similar optical
luminosity.   Both figures 6 and 7 show that this sample
probes the type of galaxies that one obtains when looking at
objects detected by {\sl GALEX} several magnitudes deeper than the SDSS spectroscopic limit.

\section{Derived Galaxy Parameters}
  We derive the following galaxy parameters according to the approach of
   \citet{salim05}: the V-band  dust attenuations, A$_{V}$ in
   magnitudes, stellar metallicity Z, the current star formation
   rate, SFR, averaged over the past 100 Myr in M$_{\odot}$ yr$^{-1}$, the
   present-day stellar mass, M$_*$, of the galaxy in
   M$_{\odot}$, the
   fraction of stellar mass formed in bursts over the last
   100 Myr, F$_{burst}$, and the \citet{scalo} b parameter,
   defined as the ratio of the current SFR
  to the past time-averaged SFR (averaged over the estimated
  age, not Hubble time). 
The galaxy parameters are derived from model libraries of
 galaxies at redshifts between 0.1 and 1.6 at redshift
 increments of 0.1 for the DEIMOS sample, and at redshifts of
 0.05,0.1,0.15,0.2, and 0.25 for the IR1.1/DR2 sample.   
Each library consists of up to $\sim$10$^5$ models. Each model is
 parameterized according to galaxy age, optical depth, star formation history (SFH), and metallicity. 
 The SFH of each model is parameterized according to \citet{kauf03a}, with an underlying, continuous, exponentially
 declining SFR upon which bursts of star formation, random in time and amplitude, are superimposed. 
 Dust attenuation in each model is parameterized using the
 prescription of \citet{cf00} using an effective $V$-band optical depth
 $\tau$$_{V}$ and absorption curve, $\tau\propto\lambda^{0.7}$  resulting from both giant molecular clouds and the
 diffuse ISM, with the fraction $\mu$ of  $\tau$$_{V}$
 contributed only by the diffuse ISM. The $V$-band optical
   depth from giant molecular clouds
 is taken to only affect stars younger than 10 Myr.  A description of the prior distributions of
 the model parameters is discussed in \citet{salim05}.
 
  Spectral energy distributions (SEDs) are created for each galaxy in the library
  using the population synthesis code of \citet{bc03}.   
  The model SEDs are convolved with the {\sl GALEX}  and SDSS filter
  response curves. Statistical estimates of physical galaxy parameters  are derived by
  comparing the observed 7 band GALEX$/$SDSS  fluxes of each
  galaxy to all the convolved model SEDs in the nearest redshift
  library. Probability density functions  (PDFs) for each
  physical parameter are created by assigning weights to
  the parameters. The $\chi^{2}$ goodness of fit of the
  model determines the weight ($\propto \exp[-\chi^{2}/2]$)
  that is assigned to the parameters of that model. The median
  (or most typical parameter value)
  of the PDF is taken as the estimate of the galaxy parameter.

  We perform SED fits for galaxies not identified as QSOs.
  In Table 2 we list the galaxies, their redshifts, $NUV-{\it
    r}$ non k-corrected colors, and their derived parameters.
  The parameters derived from the SED fits are not used unless
  the reduced $\chi^{2}$ fit of at least one model is below
  10. Only two of the 46 galaxies do not meet this
  criterion. Figures 8 and 9 show the derived parameters for the DEIMOS
  sample of galaxies overplotted on the IR1.1/DR2 sample
  plotted as shaded contour plots. The contours are labeled
  and encompass 52\%, 84\%, and 97\% of the data.  The
  diamonds as before are objects with $z<0.25$, and the crosses are objects
  with $z>0.25$. The objects
  span the range of derived parameters of the IR1.1/DR2
  sample for Z, and A$_{V}$. 

The derived metallicities, $Z = [Fe/H]$, for the sample are
not very well constrained.  The typical error, $1/4$ of the
difference between the 97.5  percentile and the 2.5 percentile
of the PDF, of the derived  metallicities are $\pm$43\%
Z$_{\odot}$. The derived M$_*$ range  from 10$^8$ to 10$^{11}$
M$_{\odot}$ with an average error of $\sim\pm0.2$dex, but the majority of  the M$_*$ are at or below
10$^{10}$M$_{\odot}$, about an order  of magnitude below the
mean of the IR1.1/DR2 sample. The derived SFRs lie between $10^{-1}$ and
  $10^{2}$ M$_{\odot}yr^{-1}$, with five galaxies having derived SFRs
greater than 10 M$_{\odot}yr^{-1}$. The average error on the
derived SFRs is typically $\pm0.3$dex. Comparing the SFRs between galaxies with
similar $NUV-r$ color or M$_{*}$, the galaxies at 
$z>0.25$ have  SFRs about an order of magnitude greater than
the galaxies at $z<0.25$.  This is mostly a selection effect
since the more distant galaxies are also more luminous on
average. A comparison of these SED derived SFRs
with H$\alpha$ derived SFRs cannot be performed since the
spectra are not flux calibrated and many lack H$\alpha$. The range of SFRs we find for
this sample is similar to the range of SFRs for local UV
galaxies found by \citet{sullivan} derived both from H$\alpha$
luminosity and from the UV luminosity, but is an order of
magnitude larger than the SFRs found by \citet{kk04} in their
sample of Goods-North Treasury Keck Redshift Survey galaxies at
$0.4<z<.9$ when compared to galaxies in the DEIMOS sample with similar redshifts.

 We find that the  galaxies are largely starbursts with $\log
 b \sim -0.1$. This is shown in Figure 10. The distribution of
 $\log b$ for objects in the DEIMOS sample with SED fits are
 plotted with the distribution for the IR1.1/DR2 sample
 normalized to the number of objects in the DEIMOS sample with
 SED fits. The DEIMOS sample shows a highly peaked value of
 $\log b$ compared to the IR1.1/DR2 sample which has two small
 peaks near $\log b \sim -3$ and $\sim -1$, showing that most
 of these galaxies currently have less star-formation now than
 in the past. The value $\log b \sim -0.1$ that describes most of
 the galaxies in the DEIMOS sample indicates that while these
 galaxies have less star formation than in the past, they are
 currently or have recently gone through a burst of
 star-formation. Two of the galaxies have $\log b > 0$
 indicating that these galaxies are going through major
 starbursts in their histories. From a statistical
 standpoint, the SED fits reveal with a 95\% reliability that at least three galaxies
 have not had a burst of starformation
    in the last 100 Myr. Half of the galaxies in the sample
    could have formed as much as $\sim$ 10\% of their stellar mass in
    a burst within the last 100 Myr, 10 could have formed up
    to $\sim 25\%$ and five of the less massive
    systems could have had up to
    $\sim 50\%$ of their stellar mass form in  bursts within
    the last 100Myr.

\section{Discussion}
In the plot of $NUV-r $  vs $M_r$ our sample lies along a ``blue sequence'' of star
forming galaxies and they have among the bluest colors for star forming galaxies
in the local Universe found by {\sl GALEX} and SDSS. While all
the galaxies show blue $NUV -r$ color this is not a homogenous
sample. Fifteen of the galaxies show disk structure in the SDSS
images. The remaining two-thirds of the sample fit into the
heterogeneous class of faint blue galaxies at intermediate
redshifts previously studied
by \citet{koo,guzmana,guzmanb,phillips} and \citet{guzmanLCBG} in the
Hubble Deep Field and adjoining fields, and by
\citet{hammer} and \citet{ornellas} in the Canada-France Redshift Survey.

The faint blue subset of the DEIMOS sample has similar luminosities,
SFRs and optical colors to the compact galaxies found in the
Hubble Deep Field by \citet{phillips}, though none are
``compact'' (optical half-light radii $<0^{\prime\prime}$.5.) This is a
selection effect in our sample. Galaxies with half-light radii this small
are classified as stars in SDSS and would not have been
selected for spectroscopy in our sample. In Figure 11 we plot, as filled
diamonds, the rest frame
 absolute B magnitude versus the B magnitude surface
 brightness and SFR versus velocity dispersion.
 The absolute B magnitudes are calculated from the observed
 bandpasses using the k-correction code of \citet{blanton},
 and the surface brightness are calculated with the
 half-light SDSS r petrosian radii. The velocity dispersions
 are derived from the measured linewidths of [OIII]5007
 for objects with $\ge3\sigma$ detections of
 the emission line. For
 comparison, we also plot, as stars and squares,
 the faint blue galaxy samples of \citet{koo,guzmana} and
 \citet{phillips} making the necessary corrections to our
 adopted cosmology. While our sample of
 galaxies have luminosities, SFRs, and velocity dispersions comparable with  the previous samples,
 the size selection effect separates our sample from the
 previous samples in surface brightness.

\citet{koo} and \citet{guzmanb} propose that the subset of the faint blue population of
galaxies with compact geometry (half light radii
$<0^{\prime\prime}$.5) and narrow emission lines ($\sigma<65
km s^{-1}$)  will fade to
become dwarf spheroidals by z=0, while \citet{hammer} claims
that the most luminous of these are too massive to become
dwarf spheroidals, and instead will become the bulges of
spiral galaxies. The most massive intermediate redshift
galaxies ($M_{*} \gtrsim 10^{10}$) in our sample could likely follow
this latter evolution path. Indeed, we see that the galaxies in our sample that
show extended structure at low redshifts ($z<0.3$) are among the the
most massive of the sample. 

The conclusions of \citet{koo} and \citet{guzmanb}
rest partially on the assumption that galactic winds from the last
starburst event will remove the remaining gas from these
systems and halt star formation causing them to fade
several magnitudes by z=0. Through modeling the UV
and optical broadband colors we find that
about half the galaxies in our present sample could have formed at most
$\sim10\%$ of their stellar mass in bursts within the past 100 Myr,
and only fifteen (all having $M_{*}<10^{9}$) could have formed over 20\% of their stellar
mass within the last 100 Myr. If the last starburst event
removes all remaining gas from these systems, as \citet{koo} propose, then
why did the previous starforming
events in these galaxies that produced the majority of the stars
in these systems not halt star formation? Our derived
starformation histories  of these galaxies argues that the current star formation events will
not halt starfromation in every galaxy. But what percentage of these will continue to
experience further  bursts of star formation is
unknown. However, even in the low mass systems of the present sample, residual
star formation or another burst of star formation is
ultimately unlikely to
change the evolutionary outcome proposed by \citet{koo} and
\citet{guzmanb} for the most compact low mass galaxies in our
sample. We suggest that the more extended sources of our sample are more likely to become  present-day dwarf
irregulars rather than dwarf spheroidals.

\section{Summary}
 We have presented DEIMOS spectra for
 objects detected by {\sl GALEX}  in the MIS survey with
 imaged counterparts in SDSS; a total exposure time of 30 min
 per slitmask was used.  GALEX has proven
to be a sensitive instrument for wide field galaxy surveys.
We have shown that the {\sl GALEX} Medim Imaging Survey followed up
with a 30 m integration with Keck/DEIMOS yields redshifts and
line measurements for star forming galaxies to z$\sim$~0.7, and has
yielded 4 QSO's 3 of which are previously un cataloged. 
 The matched sample 
 reaches approximately 3 magnitudes fainter in {\it r} than
the SDSS spectroscopic survey limits. The sample is not a homogenous
sample, but is indicative of the types of galaxies forming
stars out to z$\sim0.7$. We have derived physical parameters
for these galaxies from the SEDS and compared this sample to a
sample of ~50,000 SDSS galaxies with GALEX detections.

1. We find that roughly one-third of the galaxies are
starforming late type disk galaxies, four are QSOs at $z>1$,  
and the remaining galaxies are faint blue low mass
starbursts.\\  
\\
2. Approximately 3 out of 14 star formimg galaxies show emission line ratios
indicative of an AGN. A similar fraction was found by
\citet{sullivan} in their UV selected sample. \\
\\
3. The masses of the galaxies are typically lower than what
is found locally in the SDSS spectroscopic sample. The range
of $M_{*}$ for the DEIMOS sample spans from $10^8$ to
$10^{11} M_{\odot}$, whereas the median of the SDSS
spectroscopic sample is $\sim5\times10^{10}$.\\
\\
4. The SFRs of the galaxies at $z>0.25$ are roughly an order
of magnitude greater than the SFRs for the galaxies at
$z<0.25$.\\
\\
5. Besides three of the most massive, and reddest in NUV -r
color, the remaining galaxies show evidence of a
starburst in the last 100 Myr, with $\log b\sim-0.1$. Fifteen
of the galaxies in the lower mass range of
the sample could have formed more than 20\% of their
stellar mass in bursts of star formation within the last 100
Myr. \\
\\
6.  Our sample has similar velocity dispersions, SFRs, and
B luminosities to previous samples of faint blue galaxies,
though the galaxies  in our present study are 2 mag fainter in
surface brightess.

\acknowledgments

{\sl GALEX} is a NASA Small Explorer, launched in April 2003.
We gratefully acknowledge NASA's support for construction, operation,
and science analysis for the GALEX mission,
developed in cooperation with the Centre National d'Etudes Spatiales
of France and the Korean Ministry of 
Science and Technology. 

The authors wish to recognize and acknowledge the very
significant cultural role and reverence that the summit of
Mauna Kea has always had within the indigenous Hawaiian
community.  We are most fortunate to have the opportunity to
conduct observations from this mountain.
The analysis pipeline used to reduce the DEIMOS data was developed at UC Berkeley with
 support from NSF grant AST-0071048.

The authors thank an anonymous referee for extremely
helpful comments.

{\it Facilities:} \facility{GALEX}, \facility{KECK:II(DEIMOS)}

\clearpage

\begin{figure}
\plotone{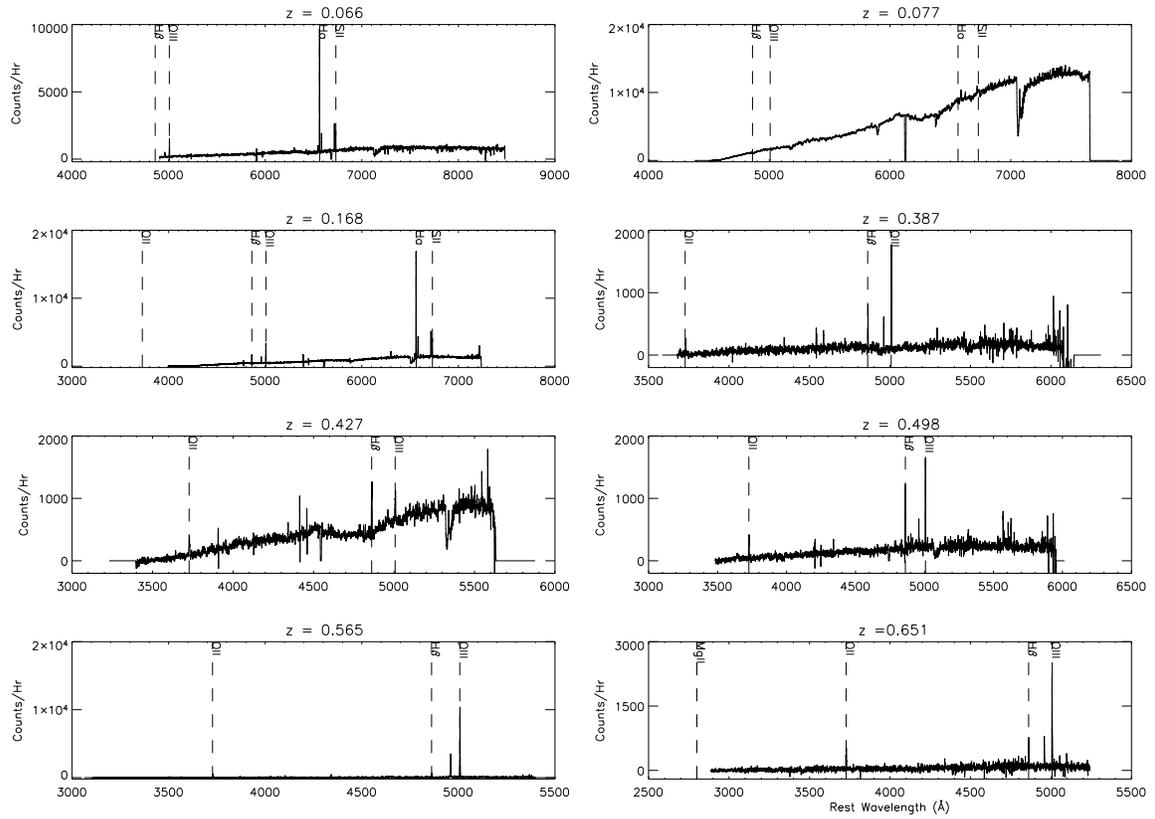}
\caption{Spectra of several galaxies from our sample at a
  range of different redshifts. All of the galaxies are blue
  sequence galaxies except the galaxy at $z=0.77$, which lies
  on the red sequence. The spectra have been boxcar smoothed by 5 pixels.}
\end{figure}
\clearpage

\begin{figure}
\plotone{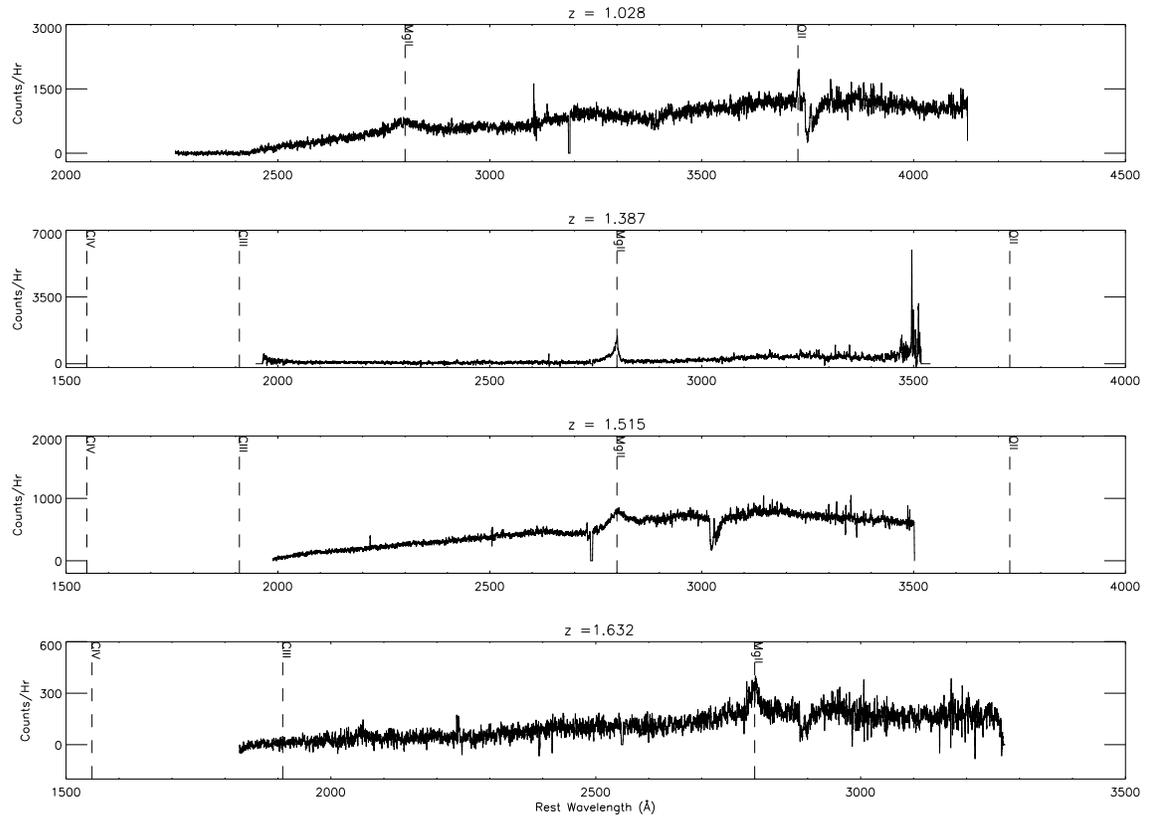}
\caption{Spectra of four QSOs from our sample. The spectra have been smoothed by 5 pixels.}
\end{figure}
\clearpage

\begin{figure}
\plotone{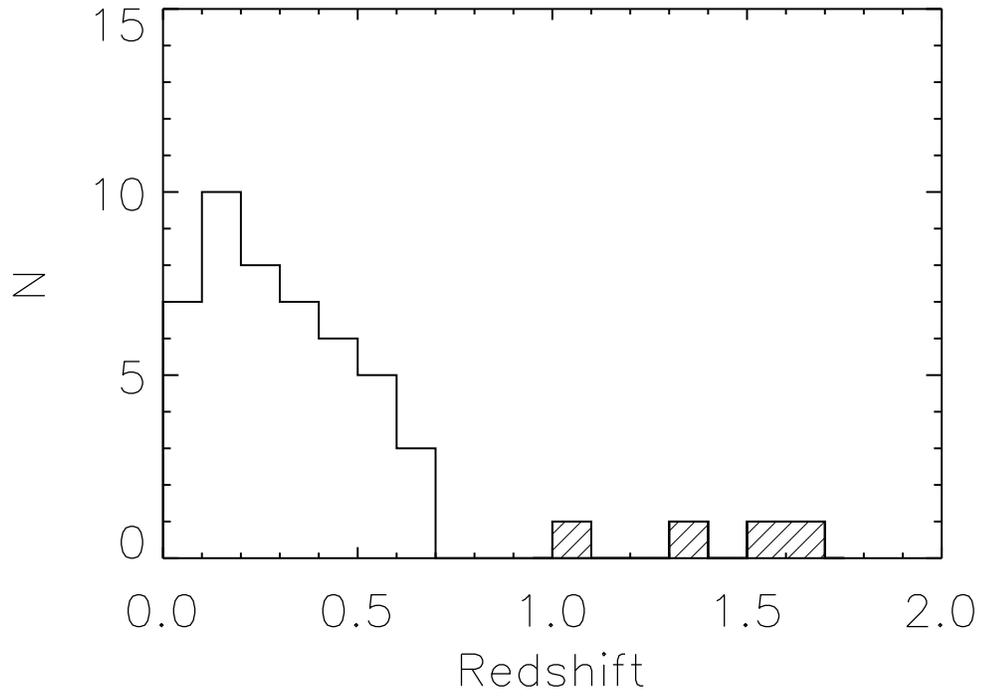}
\caption{Redshift distribution of spectroscopic sample. The mean redshift for the
  sample is $z=0.421$.  Shaded boxes indicate the quasar redshifts.}    
\end{figure}
\clearpage

\begin{figure}
\plotone{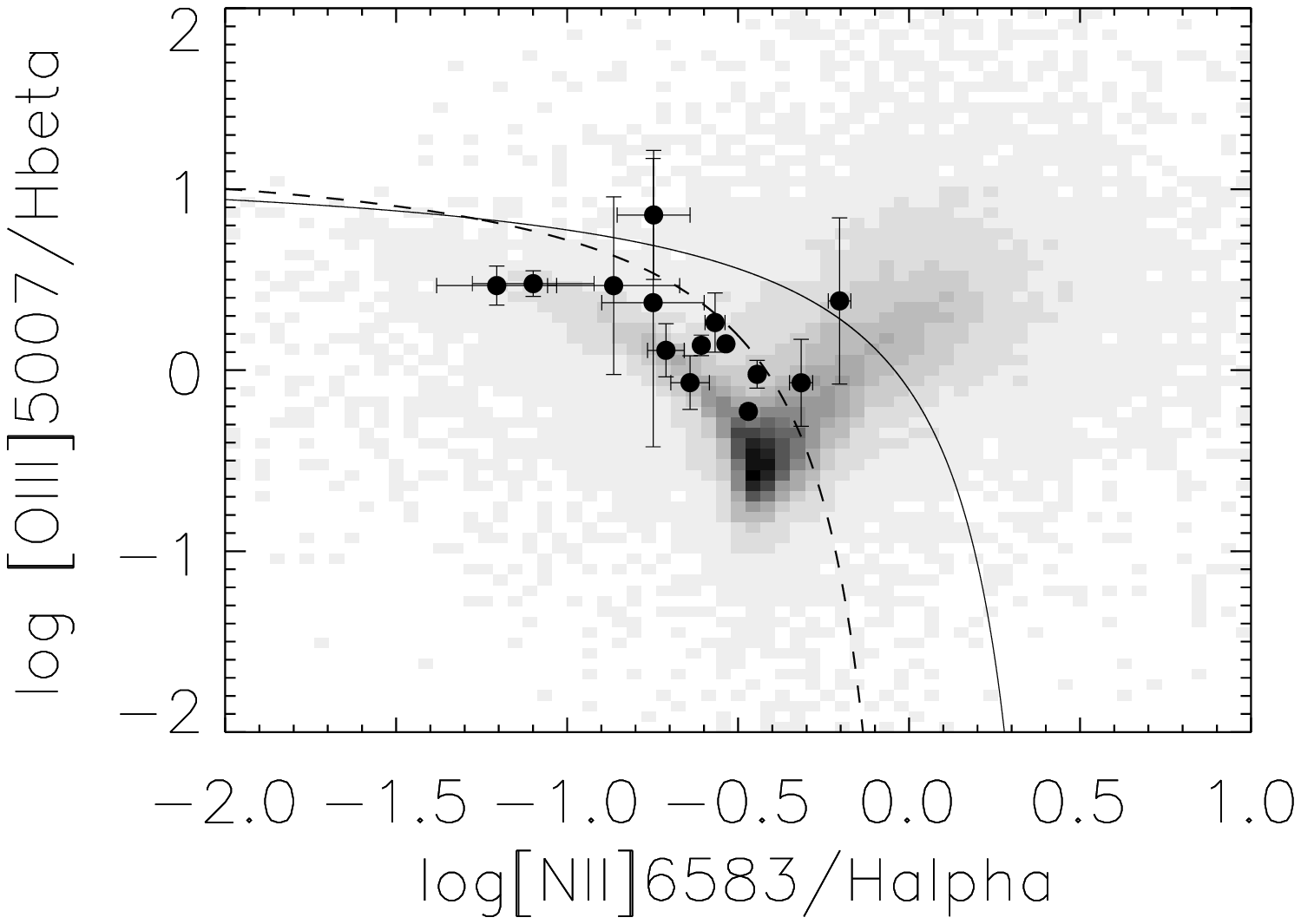}
\caption{Emission line diagnostic diagram first used by
  \citet{bpt}. The shaded 2D-histogram corresponds a to the  GALEX
    IR1.1/SDSS DR2 spectroscopic sample. The dashed curves taken from
  \citet{kew,kauf03a} show the distinction between sources
  with emission line flux coming from AGN (above) and sources
  with emission line flux coming from HII regions (below). Out
  of the 14 objects from our DEIMOS sample with detections of
  H$\beta$, [OIII], H$\alpha$, and [NII], only three have
  emission line ratios consistent with emission due to an AGN.}
\end{figure}
\clearpage

\begin{figure}
\plotone{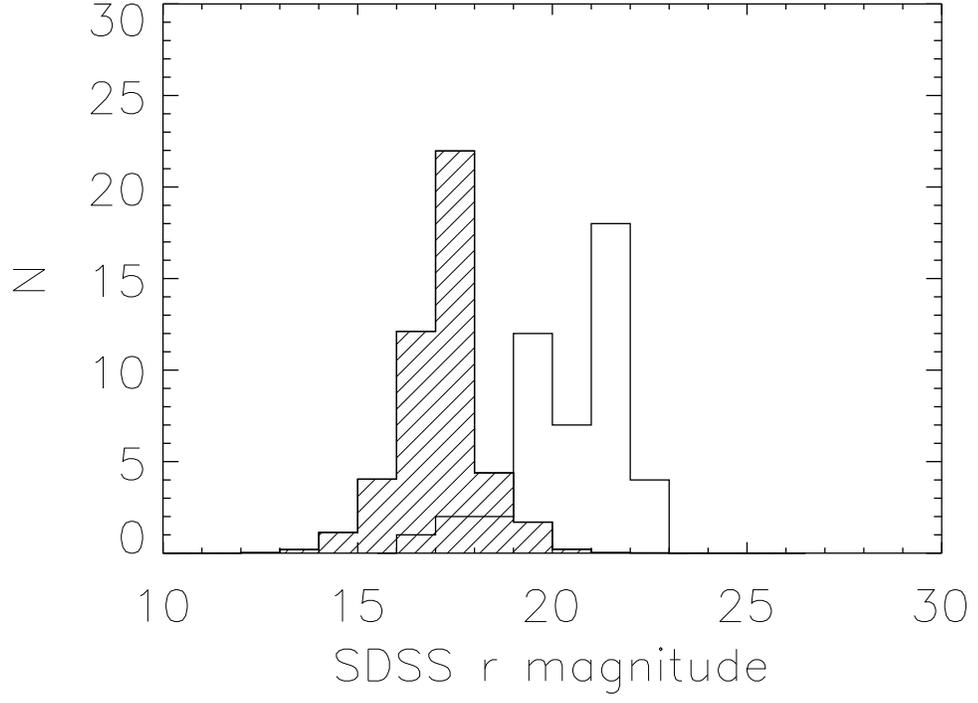}
\caption{SDSS {\it r} magnitude distribution of the
  DEIMOS  spectroscopic sample plotted with the SDSS
          {\it r} magnitude distribution of the matched GALEX
    IR1.1/SDSS DR2 spectroscopic sample (shaded histogram).
  The DEIMOS matched spectroscopic sample mean r magnitude (~21) is 3
magnitudes fainter than the mean of the SDSS spectroscopic
sample. }  
\end{figure}
\clearpage

\begin{figure}
\plotone{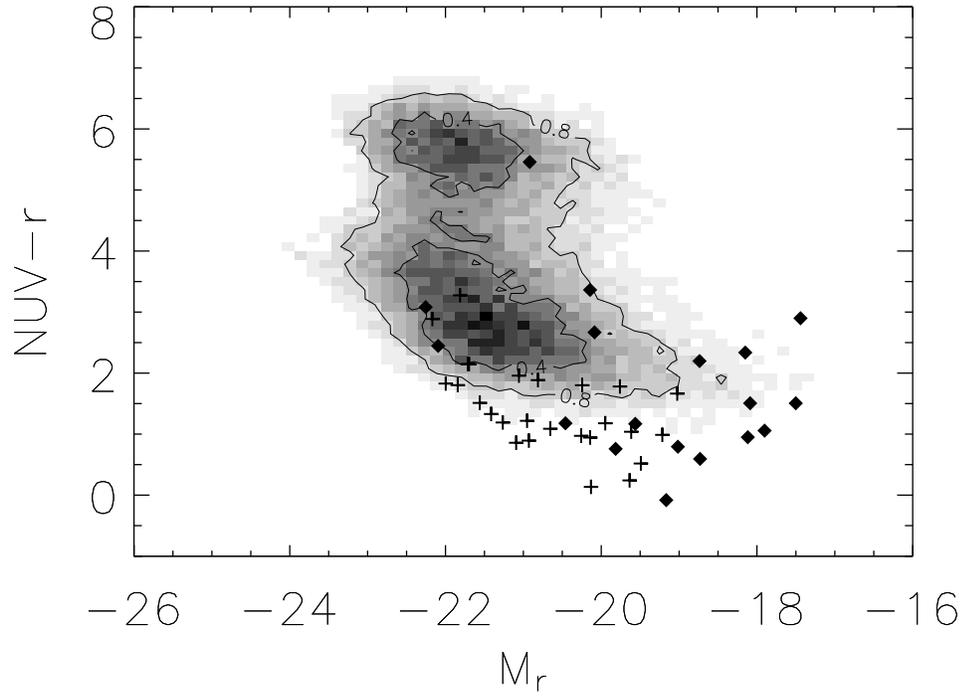}
\caption{CMD of M$_{r}$ vs. NUV-r plotted for the sample of
  GALEX/SDSS objects with DEIMOS spectroscopy plotted with the
  entire GALEX(IR1.1)/SDSS (DR2) matched sample having SDSS
  spectroscopy. The diamonds represent galaxies in the DEIMOS
  sample at $z<0.25$; crosses represent galaxies in the DEIMOS
  sample at $z>0.25$. The
  IR1.1/DR2 sample is plotted as the shaded contour plot; the
  darker regions correspond to a higher density of points and
  the contours encompass 40\% and 80\% of the  of the objects in the sample.}
\end{figure}
\clearpage

\begin{figure}
\epsfig{file=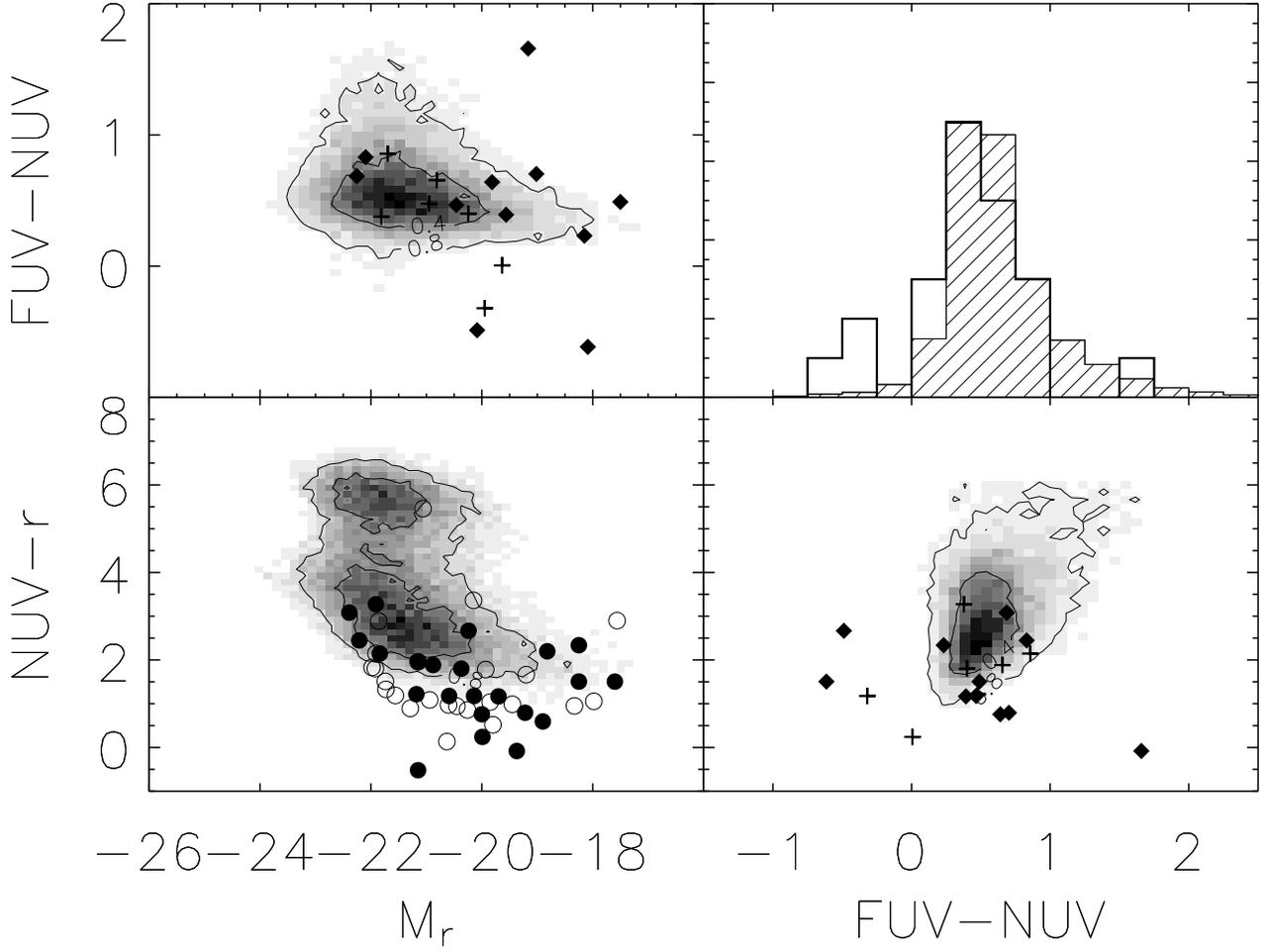, scale=1.}
\setlength{\abovecaptionskip}{50pt}
\caption{Panel 1: CMD of M$_{r}$ vs. $FUV-NUV$ plotted for the
  24 galaxies in the DEIMOS sample with $FUV$ detections,
  plotted with objects in the IR1.1/DR2 matched sample with
  $FUV$ detections shown as a shaded contour plot. PANEL 2:
  histogram of $FUV-NUV$ color for the 24 DEIMOS $FUV$
  galaxies plotted with distribution of the IR1.1/DR2 FUV
  sample (shaded histogram). PANEL 3: same as figure 5. The
  filled (unfilled) circles correspond to objects with
  (without) $FUV$ detections. Panel 4: Color-Color diagram of
  the $FUV$ DEIMOS galaxies again overplotted onto the $FUV$
  IR1.1/DEIMOS sample. The symbols used are the same as in
  figure 6 and the contours encompass 40\% and 80\% of the  of the objects in the IR1.1/DR2 sample.}
\end{figure}
\clearpage

\begin{figure}
\epsfig{file=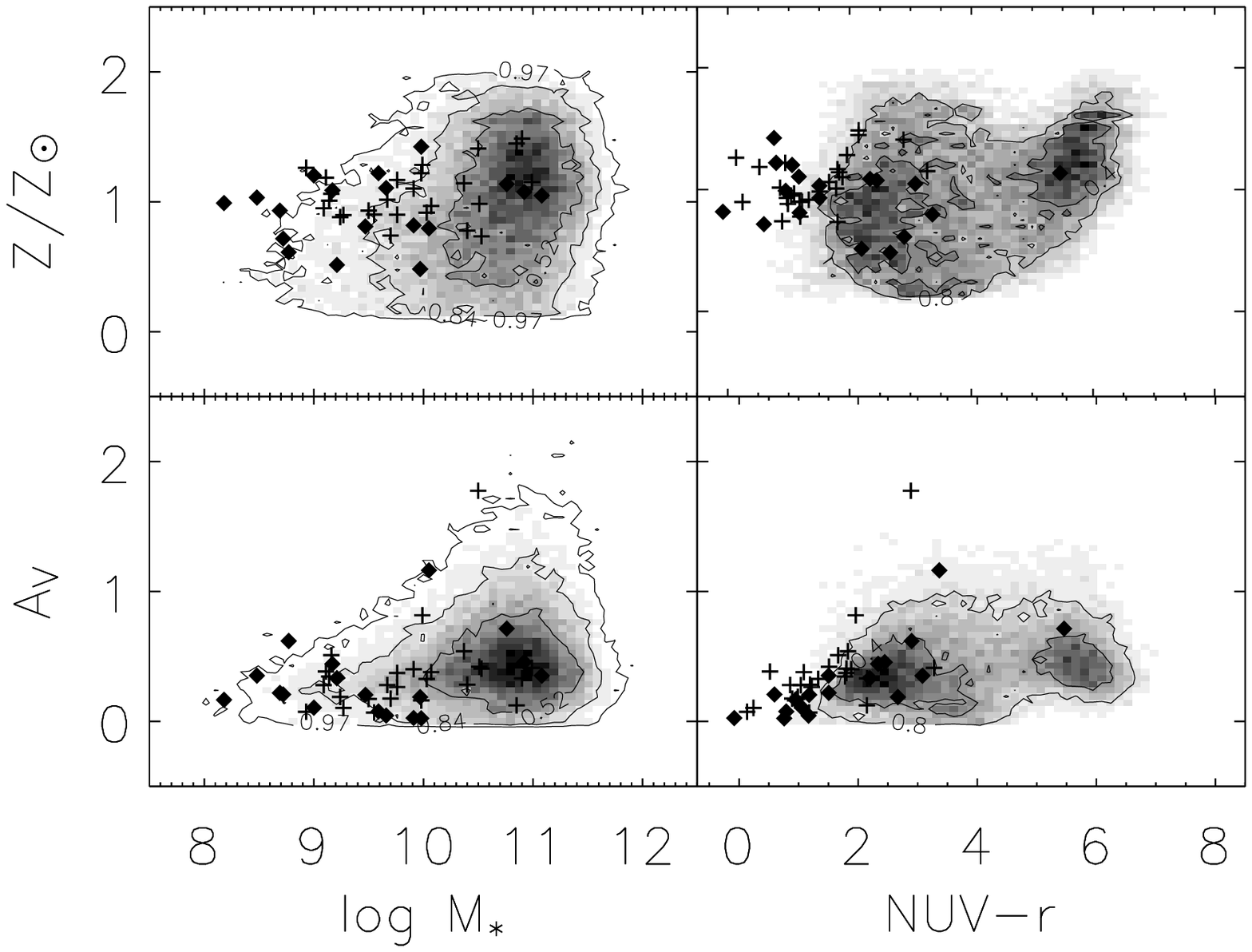, scale=1.}
\setlength{\abovecaptionskip}{50pt}
\caption{Derived Galaxy parameters, metallicity and  V-band
  attenuation plotted versus M$_{*}$ and $NUV - r$ color. The
  DEIMOS spectroscopic sample is plotted over the  matched
  {\sl GALEX}  IR1.1/SDSS DR2 spectroscopic sample (shaded
  contour plot). The crosses correspond to objects with s
  redshifts $z>0.25$. The diamonds correspond to objects with
  redshifts $z<0.25$. The contours for the plots of stellar
  mass contain 57\%, 84\%, and 97\% of the data. The contours
  of $NUV-r$ color enclose 40\% and 80\% of the sample.}
\end{figure}
\clearpage

\begin{figure}
\epsfig{file=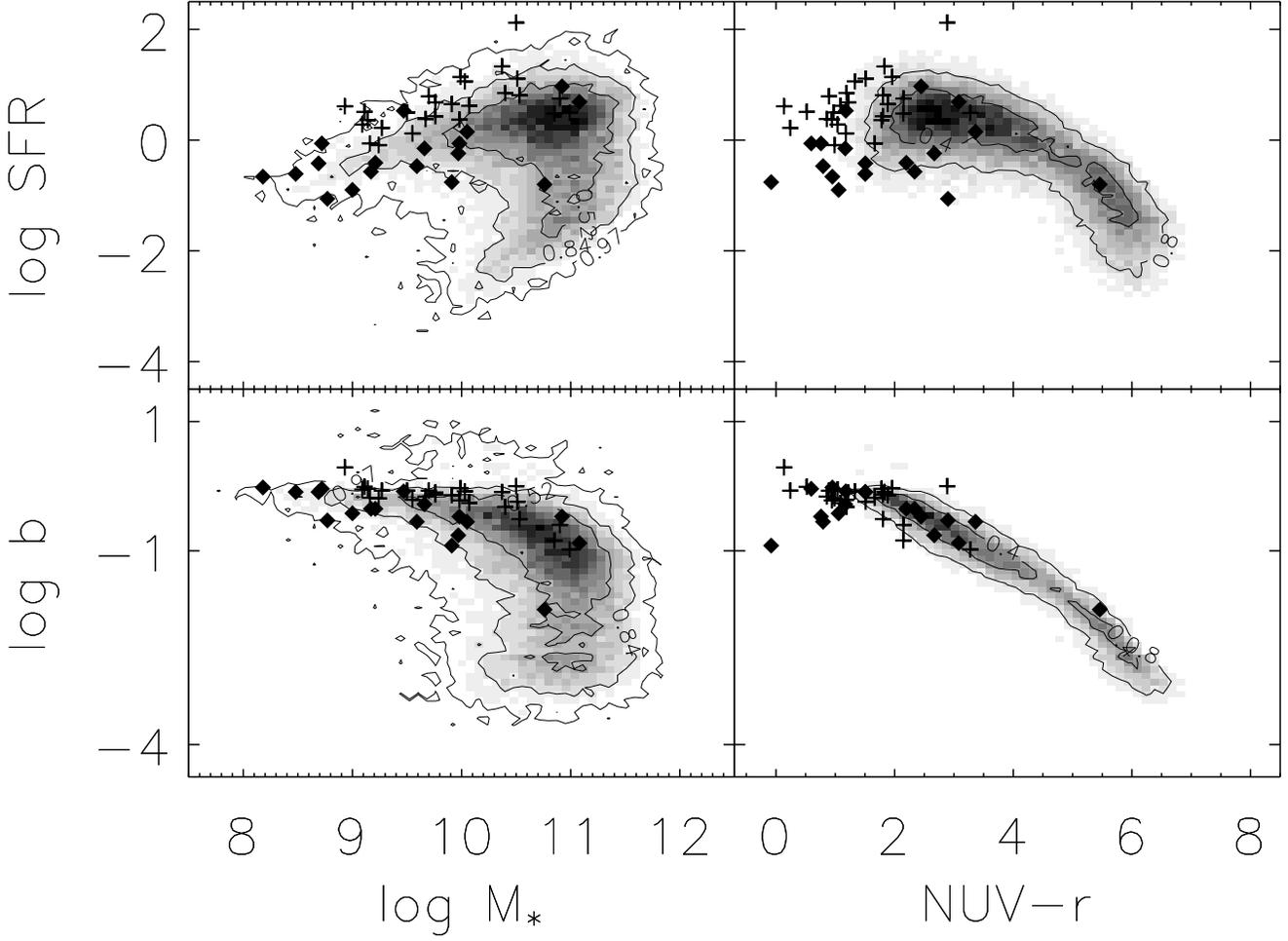, scale=1.}
\setlength{\abovecaptionskip}{50pt}
\caption{Derived Galaxy parameters, $\log$ SFR and $\log$ b
  plotted versus  M$_{*}$ and $NUV-r$ color. The DEIMOS
  spectroscopic sample is plotted over the  matched {\sl
    GALEX}  IR1.1/SDSS DR2 spectroscopic sample (shaded
  contour plot).  The crosses correspond to objects with
  redshifts $z>0.25$. The diamonds correspond to objects with
  redshifts $z<0.25$. The contours for the plots of M$_{*}$
  contain 56\%, 84\%, and 97\% of the data. The contours of
  $NUV -r$ color enclose 40\% and 80\% of the data. The SFRs
  for the high z galaxies are approximately and order of
  magnitude higher than the SFRS of the low z galaxies.}
\end{figure}
\clearpage

\begin{figure}
\plotone{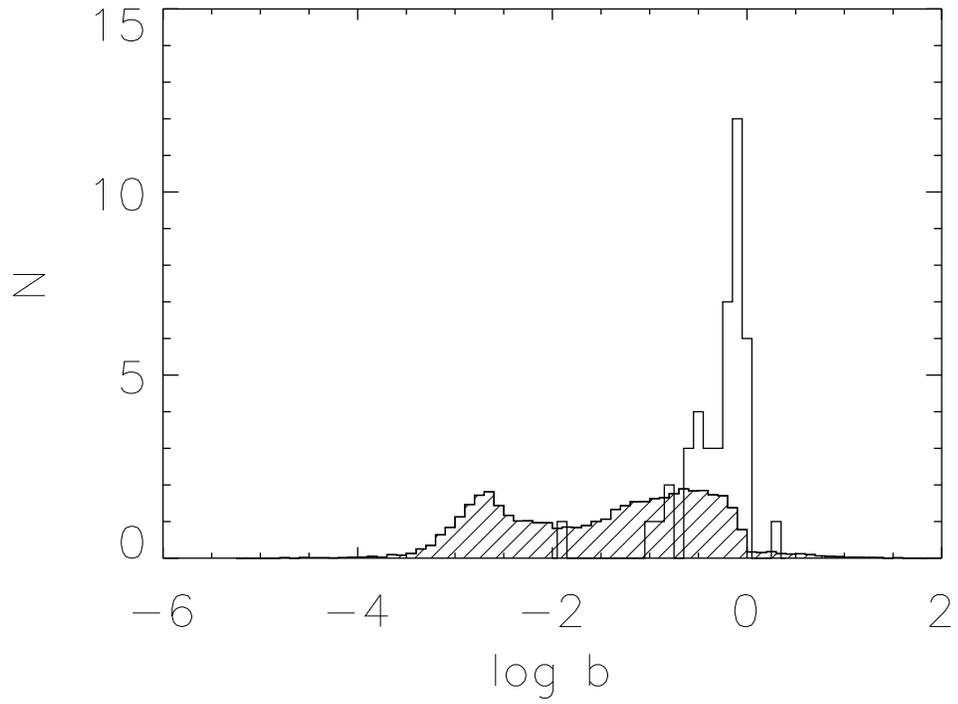}
\caption{$\log b$ histogram distribution of the  DEIMOS sample
  plotted with the $\log b$ distribution of the matched GALEX
    IR1.1/SDSS DR2 spectroscopic sample (shaded histogram)
    normalized to the size of the DEIMOS sample. This figure
    reflects our predominant sensitivity to the blue sequence
    star forming
    galaxies. }
\end{figure}
\clearpage

\begin{figure}
\epsfig{file=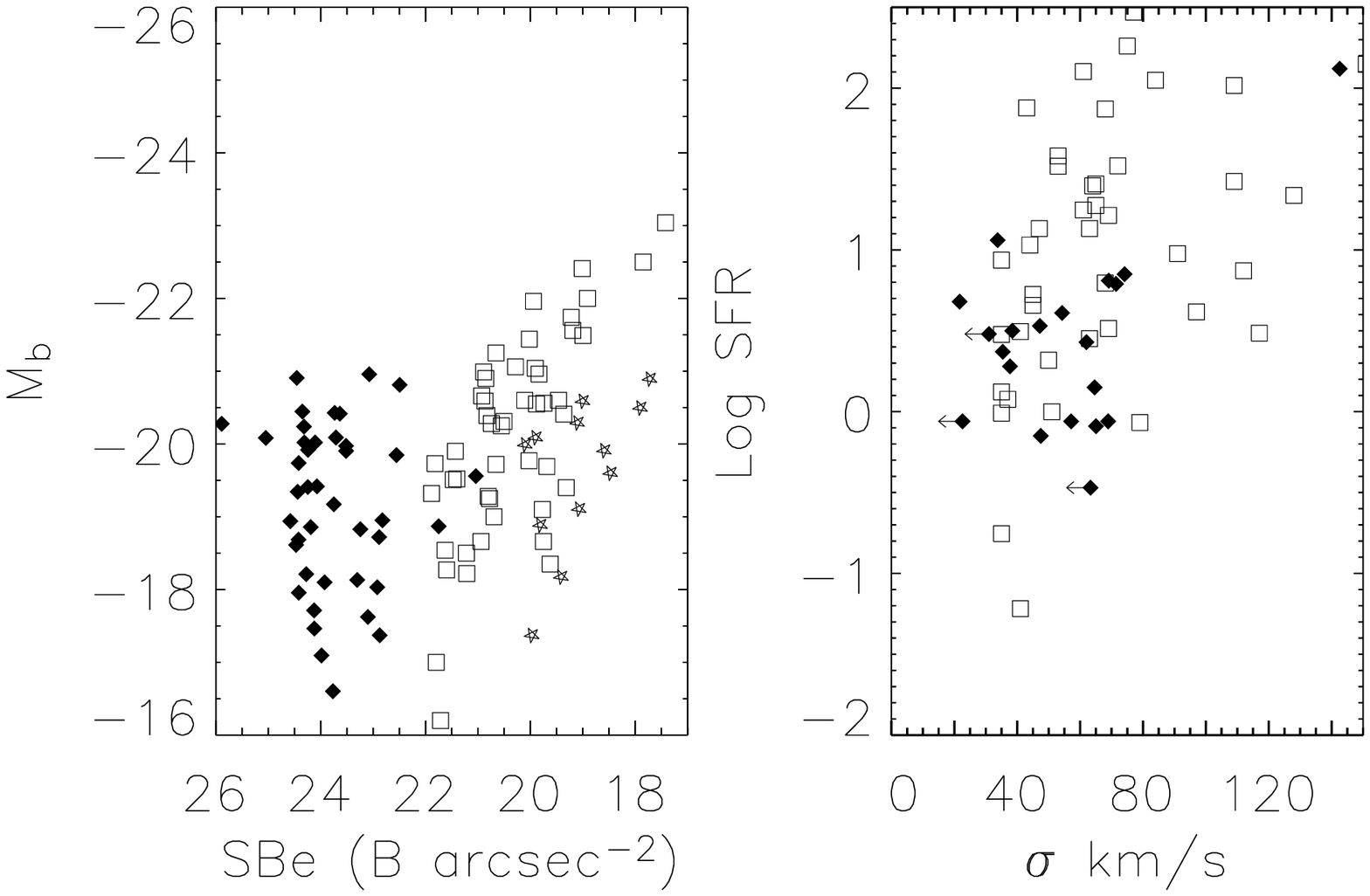, scale=1}
\setlength{\abovecaptionskip}{50pt}
\caption{Panel 1: Rest frame absolute B magnitude, M$_b$ plotted versus the
  B-mag surface brightness within the petrosian r half$-$light
  radius. Panel 2: Star formation rate plotted versus measured
  velocity dispersions. The filled diamonds correspond to objects in
  our sample. Objects taken from \citet{koo} \& \citet{guzmana} are
  plotted as stars, and objects taken from \citet{phillips}
  are plotted as squares. While the present sample has
  SFRs, luminosities, and velocity dispersions similar to the previous
  studies of faint blue galaxies, our sample contains sources
  with more extended emission and thus has lower surface brightness.}
\end{figure}
\clearpage

\begin{deluxetable}{lcccccccc}
\tabletypesize{\scriptsize}
\setlength{\tabcolsep}{0.02in}
\tablecaption{}
\tablecolumns{9}
\tablehead{
\colhead{SDSS Object ID} &\colhead{z}
&\colhead{EW$_{[OII]3727}$}  &\colhead{EW$_{H\beta}$}
&\colhead{EW$_{[OIII]5007}$}  &\colhead{EW$_{H\alpha}$}
&\colhead{EW$_{[NII]6584}$}  &\colhead{[OIII]5007/$H\beta$}  &\colhead{[NII]6584/H$\alpha$} }

\startdata
587725578037494738  &0.474  & 7.23 $\pm$ 5.29  &2.11 $\pm$ 0.51  &2.54 $\pm$ 0.29  &  *** $\pm$   ***  & *** $\pm$   ***  & 2.83 $\pm$ 0.72  & *** $\pm$  ***\\
587725578037494286  &0.134  &  *** $\pm$  ***  &2.05 $\pm$ 0.20  &2.57 $\pm$ 0.22  & 4.51 $\pm$  0.05  &3.46 $\pm$  0.13  & 1.37 $\pm$ 0.18  &0.25 $\pm$ 0.01\\
587725591459201372  &0.615  & 4.63 $\pm$ 1.24  &3.02 $\pm$ 0.50  &3.74 $\pm$ 0.33  &  *** $\pm$   ***  & *** $\pm$   ***  & 1.46 $\pm$ 0.28  & *** $\pm$  ***\\
587725578037495014  &0.213  &  *** $\pm$  ***  &6.16 $\pm$ 4.88  &4.83 $\pm$ 4.78  &58.32 $\pm$ 30.70  &3.91 $\pm$ 16.30  & 1.12 $\pm$ 1.38  &0.98 $\pm$ 0.60\\
587725578037494283  &0.084  &  *** $\pm$  ***  &0.82 $\pm$ 0.25  &1.17 $\pm$ 0.25  & 2.36 $\pm$  0.05  &1.68 $\pm$  0.10  & 1.84 $\pm$ 0.69  &0.27 $\pm$ 0.02\\
587725591459201726  &0.439  & 5.21 $\pm$ 0.35  &3.17 $\pm$ 0.18  &3.34 $\pm$ 0.06  &  *** $\pm$   ***  & *** $\pm$   ***  & 4.95 $\pm$ 0.30  & *** $\pm$  ***\\
587725591459201125$^b$  &0.066  &  *** $\pm$  ***  &7.74 $\pm$ 0.19  &2.74 $\pm$ 0.06  & 3.57 $\pm$  0.01  &3.13 $\pm$  0.02  & 1.40 $\pm$ 0.05  &0.29 $\pm$ 0.01\\
587725591459201562  &0.068  &  *** $\pm$  ***  &0.81 $\pm$ 0.82  &1.51 $\pm$ 0.75  & 2.66 $\pm$  0.19  &1.17 $\pm$  0.51  & 2.93 $\pm$ 3.32  &0.14 $\pm$ 0.06\\
587725591459201436  &0.066  &  *** $\pm$  ***  & *** $\pm$  ***  &2.28 $\pm$ 0.10  & 2.25 $\pm$  0.03  &1.65 $\pm$  0.12  &  *** $\pm$  ***  &0.14 $\pm$ 0.010\\
587725591459201670  &0.651  & 6.22 $\pm$ 0.58  &3.66 $\pm$ 0.77  &5.06 $\pm$ 0.21  &  *** $\pm$   ***  & *** $\pm$   ***  & 3.98 $\pm$ 0.84  & *** $\pm$  ***\\
587725591459201656  &0.157  &  *** $\pm$  ***  &2.04 $\pm$ 0.51  &2.10 $\pm$ 0.11  & 2.53 $\pm$  0.12  &1.60 $\pm$  1.65  & 5.81 $\pm$ 1.47  &0.02 $\pm$ 0.02\\
587725591459201666  &0.078  &  *** $\pm$  ***  &1.03 $\pm$ 1.80  &1.08 $\pm$ 0.62  & 2.10 $\pm$  0.11  &1.50 $\pm$  0.51  & 2.36 $\pm$ 4.33  &0.18 $\pm$ 0.06\\
587725591459201614  &0.188  &  *** $\pm$  ***  &2.62 $\pm$ 0.40  &2.69 $\pm$ 0.15  & 3.38 $\pm$  0.10  &2.02 $\pm$  0.83  & 3.01 $\pm$ 0.49  &0.08 $\pm$ 0.03\\
587725591459136404  &0.220  &  *** $\pm$  ***  &0.74 $\pm$ 0.59  &3.43 $\pm$ 0.57  & 4.95 $\pm$  0.23  &2.77 $\pm$  0.67  & 7.21 $\pm$ 5.93  &0.18 $\pm$ 0.04\\
587725591459136351  &0.319  & 1.34 $\pm$ 1.16  &2.38 $\pm$ 0.72  &0.76 $\pm$ 0.45  &  *** $\pm$   ***  & *** $\pm$   ***  & 0.34 $\pm$ 0.23  & *** $\pm$  ***\\
587725591459136031  &0.192  &  *** $\pm$  ***  &0.25 $\pm$ 0.22  &0.59 $\pm$ 0.34  & 2.98 $\pm$  0.12  &2.79 $\pm$  0.18  & 2.42 $\pm$ 2.55  &0.63 $\pm$ 0.05\\
587725591459136024$^b$  &0.245  &  *** $\pm$  ***  &2.00 $\pm$ 0.44  &1.81 $\pm$ 0.47  & 3.87 $\pm$  0.12  &2.54 $\pm$  0.32  & 0.85 $\pm$ 0.29  &0.23 $\pm$ 0.03\\
587725591459136330  &0.565  & 8.01 $\pm$ 0.95  &3.13 $\pm$ 0.32  &3.43 $\pm$ 0.11  &  *** $\pm$   ***  & *** $\pm$   ***  &18.69 $\pm$ 1.78  & *** $\pm$  ***\\
587725591459135913  &0.293  &  *** $\pm$  ***  &2.32 $\pm$ 0.61  &2.23 $\pm$ 0.47  & 2.76 $\pm$  0.09  &2.18 $\pm$  0.26  & 1.29 $\pm$ 0.44  &0.19 $\pm$ 0.02\\
587725591459136277  &0.290  &  *** $\pm$  ***  & *** $\pm$  ***  &1.81 $\pm$ 1.07  & 0.82 $\pm$  0.44  &2.80 $\pm$  0.41  &  *** $\pm$  ***  &3.97 $\pm$ 2.21\\
587725591459136339  &0.188  &  *** $\pm$  ***  &2.97 $\pm$ 0.35  &2.69 $\pm$ 0.35  & 5.10 $\pm$  0.07  &5.15 $\pm$  0.17  & 0.95 $\pm$ 0.17  &0.36 $\pm$ 0.01\\
587725578037166749  &0.320  &  *** $\pm$  ***  &2.30 $\pm$ 0.35  &2.45 $\pm$ 0.86  &  *** $\pm$   ***  & *** $\pm$   ***  & 0.59 $\pm$ 0.22  & *** $\pm$  ***\\
587725578037166468  &0.330  &  *** $\pm$  ***  &1.49 $\pm$ 0.20  &0.94 $\pm$ 0.17  &  *** $\pm$   ***  & *** $\pm$   ***  & 0.62 $\pm$ 0.14  & *** $\pm$  ***\\
587725578037166557  &0.168  &  *** $\pm$  ***  &2.47 $\pm$ 0.11  &3.06 $\pm$ 0.07  & 4.11 $\pm$  0.03  &3.06 $\pm$  0.09  & 2.39 $\pm$ 0.12  &0.20 $\pm$ 0.01\\
587725578037166705  &0.192  &  *** $\pm$  ***  &1.54 $\pm$ 0.35  &2.71 $\pm$ 0.29  & 2.50 $\pm$  0.11  &1.24 $\pm$  0.50  & 2.94 $\pm$ 0.73  &0.06 $\pm$ 0.03\\
587725578037166672  &0.189  &  *** $\pm$  ***  &3.85 $\pm$ 0.15  &2.92 $\pm$ 0.20  & 5.48 $\pm$  0.03  &5.96 $\pm$  0.07  & 0.59 $\pm$ 0.05  &0.34 $\pm$ 0.01\\
587725578037166295  &0.560  & 7.13 $\pm$ 0.52  &5.94 $\pm$ 0.38  &5.37 $\pm$ 0.29  &  *** $\pm$   ***  & *** $\pm$   ***  & 1.29 $\pm$ 0.11  & *** $\pm$  ***\\
587725578037166628  &0.652  & 0.95 $\pm$ 0.65  &3.54 $\pm$ 0.43  &0.86 $\pm$ 0.45  &  *** $\pm$   ***  & *** $\pm$   ***  & 0.17 $\pm$ 0.09  & *** $\pm$  ***\\
587725578037166629  &0.321  & 2.15 $\pm$ 4.43  &2.93 $\pm$ 0.17  &3.21 $\pm$ 0.07  &  *** $\pm$   ***  & *** $\pm$   ***  & 3.02 $\pm$ 0.19  & *** $\pm$  ***\\
587725578037100839  &0.561  & 5.25 $\pm$ 0.34  &2.84 $\pm$ 0.15  &4.40 $\pm$ 0.25  &  *** $\pm$   ***  & *** $\pm$   ***  & 1.70 $\pm$ 0.13  & *** $\pm$  ***\\
587725578037101267  &0.293  &  *** $\pm$  ***  &7.99 $\pm$ 8.21  &6.65 $\pm$ 0.68  & 2.60 $\pm$  1.68  &0.17 $\pm$  0.51  &  *** $\pm$  ***  &0.10 $\pm$ 0.30\\
587725578037101221  &0.518  & 5.79 $\pm$ 0.95  &3.31 $\pm$ 0.35  & *** $\pm$  ***  &  *** $\pm$   ***  & *** $\pm$   ***  &  *** $\pm$  ***  & *** $\pm$  ***\\
587725578037100922  &0.168  &  *** $\pm$  ***  &1.24 $\pm$ 0.44  &1.06 $\pm$ 0.45  & 3.54 $\pm$  0.11  &2.82 $\pm$  0.20  & 0.85 $\pm$ 0.47  &0.48 $\pm$ 0.04\\
587725578037101176  &0.281  &  *** $\pm$  ***  &2.28 $\pm$ 0.29  &3.52 $\pm$ 0.27  & 2.49 $\pm$  0.08  & *** $\pm$   ***  & 2.07 $\pm$ 0.31  & *** $\pm$  ***\\
587725591458873892  &0.281  &  *** $\pm$  ***  &2.49 $\pm$ 0.21  &3.14 $\pm$ 0.26  &  *** $\pm$   ***  & *** $\pm$   ***  & 1.21 $\pm$ 0.14  & *** $\pm$  ***\\
587725578037100775$^a$  &0.176  &  *** $\pm$  ***  & *** $\pm$  ***  & *** $\pm$  ***  & 0.86 $\pm$  0.11  &1.07 $\pm$  0.08  &  *** $\pm$  ***  &1.30 $\pm$ 0.20\\
587725591458808530  &0.427  & 2.53 $\pm$ 0.96  &3.58 $\pm$ 0.27  &3.92 $\pm$ 0.38  &  *** $\pm$   ***  & *** $\pm$   ***  & 0.94 $\pm$ 0.11  & *** $\pm$  ***\\
587725591458808528  &0.598  & 5.42 $\pm$ 0.31  &3.57 $\pm$ 0.20  & *** $\pm$  ***  &  *** $\pm$   ***  & *** $\pm$   ***  &  *** $\pm$  ***  & *** $\pm$  ***\\
587725578037101077  &0.387  & 6.51 $\pm$ 1.41  &3.96 $\pm$ 0.33  &3.12 $\pm$ 0.13  &  *** $\pm$   ***  & *** $\pm$   ***  & 2.40 $\pm$ 0.22  & *** $\pm$  ***\\
587725578037101473  &0.496  & 4.81 $\pm$ 1.31  &2.44 $\pm$ 0.52  &2.95 $\pm$ 0.13  &  *** $\pm$   ***  & *** $\pm$   ***  & 4.70 $\pm$ 1.01  & *** $\pm$  ***\\
587725578037166675  &0.498  & 7.79 $\pm$ 0.92  &3.15 $\pm$ 0.23  &3.32 $\pm$ 0.14  &  *** $\pm$   ***  & *** $\pm$   ***  & 1.39 $\pm$ 0.11  & *** $\pm$  ***
\enddata
\tablenotetext{a}{denotes objects with corresponding spectra in SDSS}
\tablenotetext{b}{denotes objects with failed SED fit}

\end{deluxetable}

\begin{deluxetable}{lccllcclll}
\tabletypesize{\scriptsize}
\tablecaption{}
\tablewidth{0pc}
\tablecolumns{10}
\tablehead{
\colhead{SDSS Object ID} &\colhead{z}
  &\colhead{RA}   &\colhead{Dec}
 &\colhead{$FUV$}    &\colhead{$NUV$}
 &\colhead{$NUV - r$}    &\colhead{$\log M_{*}$}    &\colhead{$\log$ SFR}
 &\colhead{$\log b$}}
\startdata
587725578037166705  &0.192 &17$^h$ 37$^m$ 30.681$^s$ &57$^o$ 20$^{\prime}$  8.70$^{\prime\prime}$  &22.53  &23.11  & 1.51 & 8.69  &-0.42  &-0.09\\               
587725578037166675  &0.498 &17$^h$ 37$^m$ 34.409$^s$ &57$^o$ 21$^{\prime}$ 13.71$^{\prime\prime}$  & ***   &22.62  & 0.97 & 9.50  & 0.50  &-0.07\\
587725578037166672  &0.189 &17$^h$ 37$^m$ 39.053$^s$ &57$^o$ 21$^{\prime}$ 30.10$^{\prime\prime}$  & ***   &23.04  & 3.36 &10.05  & 0.15  &-0.55\\
587725578037166749  &0.320 &17$^h$ 37$^m$ 41.177$^s$ &57$^o$ 18$^{\prime}$ 46.29$^{\prime\prime}$  &22.87  &22.12  & 1.88 & 9.91  & 0.65  &-0.14\\
587725578037100839  &0.561 &17$^h$ 37$^m$ 43.542$^s$ &57$^o$ 24$^{\prime}$  5.51$^{\prime\prime}$  & ***   &22.19  & 1.19 &10.40  & 0.85  &-0.32\\
587725578037166468  &0.330 &17$^h$ 37$^m$ 45.923$^s$ &57$^o$ 19$^{\prime}$ 17.19$^{\prime\prime}$  &22.38  &21.50  & 2.15 &10.85  & 0.48  &-0.84\\
587725578037166557  &0.168 &17$^h$ 37$^m$ 58.250$^s$ &57$^o$ 20$^{\prime}$ 16.80$^{\prime\prime}$  &20.61  &20.12  & 1.18 & 9.47  & 0.53  &-0.08\\
587725578037166295  &0.560 &17$^h$ 37$^m$ 59.495$^s$ &57$^o$ 23$^{\prime}$ 21.80$^{\prime\prime}$  & ***   &22.16  & 0.89 & 9.70  & 0.79  &-0.07\\
587725578037166628  &0.652 &17$^h$ 38$^m$  2.622$^s$ &57$^o$ 23$^{\prime}$ 26.99$^{\prime\prime}$  & ***   &22.73  & 1.51 &10.51  & 1.11  &-0.24\\
587725578037166629  &0.321 &17$^h$ 38$^m$  7.756$^s$ &57$^o$ 23$^{\prime}$ 34.90$^{\prime\prime}$  & ***   &22.32  & 1.04 & 9.09  & 0.28  &-0.05\\
587725578037101267  &0.293 &17$^h$ 38$^m$ 22.192$^s$ &57$^o$ 25$^{\prime}$  2.82$^{\prime\prime}$  & ***   &23.36  & 1.66 & 9.16  &-0.06  &-0.18\\
587725578037100922  &0.168 &17$^h$ 38$^m$ 23.628$^s$ &57$^o$ 27$^{\prime}$ 56.42$^{\prime\prime}$  &21.48  &21.97  & 2.67 & 9.97  &-0.24  &-0.76\\
587725578037101473  &0.496 &17$^h$ 38$^m$ 27.876$^s$ &57$^o$ 26$^{\prime}$ 33.90$^{\prime\prime}$  &22.91  &22.28  & 1.22 & 9.76  & 0.68  &-0.10\\
587725578037101077  &0.387 &17$^h$ 38$^m$ 36.526$^s$ &57$^o$ 30$^{\prime}$ 38.59$^{\prime\prime}$  & ***   &22.32  & 0.52 & 9.11  & 0.51  &-0.01\\
587725578037101221  &0.518 &17$^h$ 38$^m$ 37.317$^s$ &57$^o$ 26$^{\prime}$ 53.92$^{\prime\prime}$  & ***   &22.50  & 1.09 &10.07  & 0.62  &-0.26\\
587725578037100775$^a$  &0.176 &17$^h$ 38$^m$ 37.646$^s$ &57$^o$ 29$^{\prime}$ 13.99$^{\prime\prime}$  &21.03  &20.34  & 3.08 &11.08  & 0.69  &-0.88\\
587725578037101176  &0.281 &17$^h$ 38$^m$ 41.873$^s$ &57$^o$ 28$^{\prime}$  9.30$^{\prime\prime}$  & ***   &22.33  & 0.99 & 9.24  &-0.09  &-0.19\\
587725591458873892  &0.281 &17$^h$ 38$^m$ 48.259$^s$ &57$^o$ 29$^{\prime}$ 10.40$^{\prime\prime}$  &22.69  &22.23  & 1.80 & 9.76  & 0.43  &-0.13\\
587725591458808530  &0.427 &17$^h$ 38$^m$ 50.596$^s$ &57$^o$ 29$^{\prime}$ 54.31$^{\prime\prime}$  & ***   &22.88  & 2.89 &10.50  & 2.12  & 0.00\\
587725591458808528  &0.598 &17$^h$ 38$^m$ 54.873$^s$ &57$^o$ 30$^{\prime}$  5.80$^{\prime\prime}$  & ***   &22.58  & 1.83 &10.37  & 1.33  &-0.09\\
587725578037494286  &0.134 &17$^h$ 39$^m$ 45.205$^s$ &56$^o$ 40$^{\prime}$  2.20$^{\prime\prime}$  &20.87  &20.47  & 1.17 & 9.66  &-0.15  &-0.28\\
587725578037494409  &0.355 &17$^h$ 39$^m$ 51.035$^s$ &56$^o$ 39$^{\prime}$ 13.11$^{\prime\prime}$  & ***   &23.23  & 1.78 & 9.14  & 0.36  &-0.04\\
587725578037495014  &0.213 &17$^h$ 39$^m$ 51.343$^s$ &56$^o$ 39$^{\prime}$ 58.89$^{\prime\prime}$  & ***   &22.73  & 0.95 & 8.18  &-0.66  &-0.02\\
587725578037494283  &0.084 &17$^h$ 39$^m$ 53.130$^s$ &56$^o$ 40$^{\prime}$ 18.91$^{\prime\prime}$  &22.24  &22.00  & 2.34 & 9.17  &-0.57  &-0.35\\
587725578037494050  &0.077 &17$^h$ 39$^m$ 56.243$^s$ &56$^o$ 38$^{\prime}$ 17.20$^{\prime\prime}$  & ***   &22.11  & 5.46 &10.76  &-0.80  &-1.91\\
587725578037494494  &0.356 &17$^h$ 39$^m$ 56.287$^s$ &56$^o$ 37$^{\prime}$ 21.00$^{\prime\prime}$  & ***   &22.20  & 1.96 & 9.99  & 1.14  &-0.03\\
587725578037494738  &0.474 &17$^h$ 39$^m$ 56.704$^s$ &56$^o$ 37$^{\prime}$ 53.80$^{\prime\prime}$  & ***   &22.72  & 0.86 & 9.67  & 0.38  &-0.16\\
587725591459201436  &0.066 &17$^h$ 40$^m$  1.663$^s$ &56$^o$ 44$^{\prime}$  3.12$^{\prime\prime}$  &20.89  &20.74  & 2.20 & 9.21  &-0.41  &-0.35\\
587725591459201125$^b$  &0.066 &17$^h$ 40$^m$  3.853$^s$ &56$^o$ 41$^{\prime}$ 59.82$^{\prime\prime}$  &20.37  &19.99  & 1.96 & ***   & ***   & ***\\
587725591459201656  &0.157 &17$^h$ 40$^m$  6.724$^s$ &56$^o$ 43$^{\prime}$ 49.51$^{\prime\prime}$  &22.64  &22.12  &-0.52 & 8.68  &-0.97  &-0.24\\
587725591459201627  &0.079 &17$^h$ 40$^m$  8.196$^s$ &56$^o$ 44$^{\prime}$ 57.80$^{\prime\prime}$  &22.18  &21.68  & 1.51 & 8.48  &-0.61  &-0.09\\
587725591459201372  &0.615 &17$^h$ 40$^m$  8.328$^s$ &56$^o$ 39$^{\prime}$ 39.82$^{\prime\prime}$  & ***   &22.68  & 1.80  &10.53  & 0.81  &-0.51\\
587725591459201929  &0.436 &17$^h$ 40$^m$ 15.117$^s$ &56$^o$ 46$^{\prime}$  4.70$^{\prime\prime}$  &22.36  &22.16  & 0.24 & 9.27  & 0.22  &-0.07\\
587725591459201726  &0.439 &17$^h$ 40$^m$ 17.307$^s$ &56$^o$ 41$^{\prime}$ 31.09$^{\prime\prime}$  & ***   &22.42  & 0.94 & 9.98  & 0.37  &-0.22\\
587725591459136339  &0.188 &17$^h$ 40$^m$ 17.468$^s$ &56$^o$ 48$^{\prime}$  5.50$^{\prime\prime}$  &21.23  &20.57  & 0.76 & 9.98  &-0.06  &-0.47\\
587725591459201670  &0.651 &17$^h$ 40$^m$ 19.175$^s$ &56$^o$ 43$^{\prime}$ 39.18$^{\prime\prime}$  & ***   &22.46  & 0.14 & 8.93  & 0.61  & 0.29\\
587725591459201562  &0.068 &17$^h$ 40$^m$ 20.969$^s$ &56$^o$ 42$^{\prime}$ 43.30$^{\prime\prime}$  & ***   &22.77  & 2.90 & 8.77  &-1.06  &-0.53\\
587725591459201614  &0.188 &17$^h$ 40$^m$ 23.225$^s$ &56$^o$ 45$^{\prime}$ 42.52$^{\prime\prime}$  &22.37  &21.50  & 0.59 & 8.72  &-0.06  &-0.04\\
587725591459136404  &0.220 &17$^h$ 40$^m$ 23.708$^s$ &56$^o$ 46$^{\prime}$ 34.10$^{\prime\prime}$  &22.53  &21.76  & 0.79 & 9.59  &-0.47  &-0.55\\
587725591459135913  &0.293 &17$^h$ 40$^m$ 29.685$^s$ &56$^o$ 50$^{\prime}$ 27.71$^{\prime\prime}$  &21.64  &21.93  & 1.18 & 9.55  & 0.12  &-0.21\\
587725591459136024$^b$  &0.245 &17$^h$ 40$^m$ 32.498$^s$ &56$^o$ 49$^{\prime}$ 12.51$^{\prime\prime}$  &22.73  &20.99  &-0.08 & ***   & ***   & ***\\
587725591459201666  &0.078 &17$^h$ 40$^m$ 33.193$^s$ &56$^o$ 44$^{\prime}$  5.32$^{\prime\prime}$  & ***   &20.82  & 1.06 & 9.00  &-0.90  &-0.42\\
587725591459136277  &0.290 &17$^h$ 40$^m$ 35.691$^s$ &56$^o$ 50$^{\prime}$ 29.29$^{\prime\prime}$  &22.66  &22.23  & 3.27 &10.99  & 0.49  &-0.98\\
587725591459136330  &0.565 &17$^h$ 40$^m$ 46.362$^s$ &56$^o$ 49$^{\prime}$ 13.40$^{\prime\prime}$  & ***   &22.18  & 1.33 &10.03  & 1.06  &-0.08\\
587725591459136031  &0.192 &17$^h$ 40$^m$ 56.177$^s$ &56$^o$ 49$^{\prime}$ 15.38$^{\prime\prime}$  &20.94  &20.09  & 2.45 &10.92  & 0.97  &-0.47\\
587725591459136351  &0.319 &17$^h$ 40$^m$ 56.594$^s$ &56$^o$ 48$^{\prime}$ 52.20$^{\prime\prime}$  & ***   &21.35  & 2.15 &10.90  & 0.75  &-0.60\\
\\
\\
587725578037100834  &1.515 &17$^h$ 38$^m$  9.360$^s$ &57$^o$ 25$^{\prime}$ 21.36$^{\prime\prime}$  & ***   &23.40  & 2.63 & ***   & ***   & *** \\
587725591459201473  &1.387 &17$^h$ 40$^m$  9.836$^s$ &56$^o$ 40$^{\prime}$  7.68$^{\prime\prime}$  & ***   &22.67  & 1.14 & ***   & ***   & *** \\
587725591459136108  &1.632 &17$^h$ 40$^m$ 43.682$^s$ &56$^o$ 48$^{\prime}$ 45.36$^{\prime\prime}$  & ***   &23.05  & 1.67 & ***   & ***   & *** \\
587725591459135981$^a$  &1.028 &17$^h$ 40$^m$ 49.197$^s$ &56$^o$ 47$^{\prime}$ 23.65$^{\prime\prime}$  &22.62  &21.20  & 1.94 & ***   & ***   & *** 
\enddata
\tablenotetext{a}{denotes objects with corresponding spectra in SDSS}
\tablenotetext{b}{denotes objects with failed SED fit}
\end{deluxetable}
\end{document}